\documentclass[english]{iopart}
\usepackage{lmodern}
\usepackage[T1]{fontenc}
\usepackage[utf8]{inputenc}
\usepackage{array}
\usepackage{booktabs}
\usepackage{textcomp}
\usepackage{url}
\usepackage{multirow}
\usepackage{amstext}
\usepackage{graphicx}
\usepackage{esint}

\makeatletter

\providecommand{\tabularnewline}{\\}

\usepackage{iopams}
\usepackage{setstack}

\makeatother

\usepackage{babel}
\begin{document}

\title{ Physics of non-steady state diffusion
of lightweight atoms in a heavy atom matrix. Introducing an open-source tool for simulated-experiments in fluid mechanics}

\author{Roberto Serrano-L\'opez\footnote[3]{Contact: robertosl@ubu.es}, Santiago Cuesta-L\'opez \footnote[2]{alternative mailing address: scuesta@ubu.es}, Oscar Tapia-J\'udez,
Jorge Fradera}

\address{ Science and Technology Park. University of Burgos. I+D+I Building. Room 63. Plaza Misael
Ba\~nuelos s/n, 09001, Burgos (Spain)}

\submitto{Eur. J. Phys.}

\begin{abstract}

The practice-based learning methodologies offer to undergraduate professors
different ways to illustrate certain general physic principles. Traditional
experimental workbenches have been extensively used during decades
for academic lessons in order to complete theoretical dissertations
or lectures, aiming at assuring an adequate understanding. The high
cost of materials and laboratory equipment, the excessive preparation
time, and the difficulty for carrying out off-site-campus replications
by students, are disadvantages that can discourage of trying new kinds
of experimental tasks. This paper gives insight of simulated-experiment
possibilities through an open-source-based computational suite in teaching fluid mechanics. Physics underlying diffusion of a light specie in a heavier atom matrix,
as function of time and position, were explained to students as an example to teach them the Fick's Second Law expression. 
We present a docent step-by-step programme, scheduled in
three sessions. The expected solution is firstly
explained and then compared both to a real case, by published results,
and to the virtual-classroom experiment, by resolving a differential equation with
numerical schemes. 
Students were able to make their own hypothesis, change
input parameters, and contrasting their initial assumptions.
\end{abstract}
\maketitle

\section{Introduction}

Almost every physics teacher would agree to attribute the essential
role of practice lessons in the learning process. The evolution of
physics learning approaches since early sixties of twentieth century
was extensively reviewed by Trumper~\cite{Trumper2003}, explaining
the traditional and constructivist point of view and showed the principles
and advantages of the microcomputer-based laboratories (immediate
feedback, partner collaboration, reduced cost,...) in alignment with
the American Association of Physics Teachers (AAPT) basis~\cite{AmericanAssociationofPhysicsTeachersAAPT1998}.
In the current frame of the European Higher Education Area, learning
methods have been focused as one of the critical paths to achieve
encouragement of students~\cite{Romero2013}. The design of attractive
teaching resources, although requires a initial reflection and development,
have shown in different experiences a effective improvement of motivation
and even in student results~\cite{Dewhurst1994,Meir2005,Hashemi2005,Clark2007,Meisner2008,Cuesta-Lopez2013}.
Hence, new computational simulation tools include the main characteristics
needed to achieve good teaching results.

Some considerations can be argued talking about a well-guided practice-learning
with computational applications: (i) they allow a constant student
improve due to a constant self evaluation; (ii) different individual
skills and team job strategies may be developed in the same set of
sessions; (iii) students have the possibility of emulate activities
of professional lab-assistant without the inherent risks of a real
practice; (iv) a higher amount of time can be spent in demonstration
experiments instead of lectures or theoretical formulations (learning
the meaning by applying it, instead of learning formula by reading
it); (v) the global-net offers a free source of data and fully-functional
computational codes, downloadable and installable  by students. The
latter has to be included as a main objective for practice design,
to ensure the accessibility and repeatability of sessions.

In this manuscript we present a basic course aiming at teaching the basic concepts of fluid mechanics particularizing on diffusion at the atomic level. 
Diffusion of hydrogen in metals is an interesting and complex problem of technological importance in different fields. For example, hydrogen interaction, from an engineering standpoint, frequently results in compromising the mechanical and chemical properties of materials ~\cite{fisher1999hydrogen,Das97}. Diffusion of hydrogen and its isotopes has been, and still continues being, one of the key challenges in Nuclear Fusion ~\cite{Stickney72,Hong10}.

On the other hand, hydrogen is seen as a versatile energy carrier for the future. One of the promising methods for hydrogen storage makes use of the capability of some metals and alloys to easily uptake this element. The development of efficient hydrogen storage systems requires, however, a detailed knowledge on how hydrogen diffuses in metals ~\cite{Hoogers2003}.

Although the chief aspects of hydrogen diffusion in metals are complex, and they need an understanding from quantum to crystallographic (dependence on lattice structure) effects, basic concepts from a classical point of view constitute a basic pillar to be learnt. Our approach uses a particular docent methodology that combines experimental data into computer aided fluid dynamics tools, allowing to teach and visualize physical concepts underlying the physics of non-steady state diffusion of hydrogen in a heavy atom matrix.

In addition to the docent introduction of the physical laws into an open-source computational tool-box and visualization module, we present a teaching plan organized in three sessions that combine cooperative with guided learning methodologies.

\section{Organization of docent sessions}

To achieve an effective learning for students, planning becomes a key
issue for practice-based approach. A meditated structure of sessions
must include not only the lesson material, but the pathway to accomplish
students global comprehension of concepts. This means a holistic point
of view that starts with the previous and new knowledge, and goes on
with the need of incorporate skills and attitudes that have to be
developed.

Planning have been organized in order to do not constrain the learning
environment in the main classroom, but to involve students to be free
of choosing the place to continue their work. The use of free and
open-source codes allows them to install the tools needed to carry out tasks on their own desktops, and
even using web-based resources among their class or group partners
(forums, networking, allocation of tasks).

A model of cooperative learning was used, distributing students in small groups around a computer terminal (2-3 students per post). This strategy has been shown to contribute to an overall increase in motivation and promotion of cognitive development ~\cite{Jungst2000}. Moreover, the course was carried out using a guided learning approach, where the students were guided step by step across the different exercises and problems, what was necessary to satisfy the needs of the students at each stage, at the time that ensured to keep the rhythm of overall progress of the learning experience.

In this case, the description of non-steady state diffusion problem
was the initial lesson, particularized for the case of a lightweight
atom for the diffusive specie. Attendees to the lessons were undergraduate
engineering students. Table~\ref{tab:Essentials-of-lessons} shows
the schedule of teaching sessions proposed.

\begin{table}
\noindent \begin{centering}
\begin{tabular}{ccc}
\toprule 
{\footnotesize Session} & {\footnotesize Duration} & {\footnotesize Task}\tabularnewline
\midrule
\midrule 
\multirow{3}{*}{{\footnotesize 1A- Theoretical lecture}} & \multirow{3}{*}{{\footnotesize 30 min}} & {\footnotesize Introduction to physics of diffusion}\tabularnewline
 &  & {\footnotesize The non-steady state problem: Seconds Fick's Law}\tabularnewline
 &  & {\footnotesize Solution with error function}\tabularnewline
\midrule 
{\footnotesize 1B- Example} & {\footnotesize 20 min} & {\footnotesize Example problem: hydrogen diffusion through vanadium}\tabularnewline
\midrule 
 &  & {\footnotesize Objectives description. Team-leaders choice}\tabularnewline
\cmidrule{3-3} 
\multirow{2}{*}{{\footnotesize 2- Spreadsheet}} & \multirow{2}{*}{{\footnotesize 50 min}} & {\footnotesize Inputs needed and expected results. Questions and answers}\tabularnewline
\cmidrule{3-3} 
 &  & {\footnotesize Introduction of open-source application}\tabularnewline
\cmidrule{3-3} 
 &  & {\footnotesize Individual work: approach to the new tool}\tabularnewline
\midrule 
 &  & {\footnotesize Basic mesh generation}\tabularnewline
\cmidrule{3-3} 
\multirow{2}{*}{{\footnotesize 3A- Simulation}} & \multirow{2}{*}{{\footnotesize 50 min}} & {\footnotesize Boundary and initial conditions}\tabularnewline
\cmidrule{3-3} 
 &  & {\footnotesize Schemes and control configurations}\tabularnewline
\cmidrule{3-3} 
 &  & {\footnotesize Simulation run}\tabularnewline
\midrule 
\multirow{2}{*}{{\footnotesize 3B- Validation}} & \multirow{2}{*}{{\footnotesize 50 min}} & {\footnotesize Visualization of results. Data extraction}\tabularnewline
\cmidrule{3-3} 
 &  & {\footnotesize Comparison with analytical and experimental approaches}\tabularnewline
\bottomrule
\end{tabular}
\par\end{centering}

\caption{Essentials of lessons time-table. \label{tab:Essentials-of-lessons}}
\end{table}

\section{Understanding non steady-state diffusion. Fick's second law.}

An initial session was set up as a group lecture to introduce the theoretical expressions underlying the physics of non steady-state diffusion (session 1-A).\\
Let us begin considering two species (A=diffusing and B=matrix)
that remain physically bounded by their own interface. We can assure that
concentration of the lightest one (A) is constant in its side, i.e.
concentration is time independent. At constant energy, a material
flow will be initialized through the heavier specie (B) so that the
atoms of A will pass among molecular lattice of B. This means that
concentration of A ({[}A{]}\emph{ }denoted below as \emph{c}) will
grow up inside the B matrix. This non steady-state diffusion problem
is commonly resolved by Fick's second law, which can be expressed
with a laplacian equation without convective term:

\begin{equation}
\frac{\partial c}{\partial t}-\nabla^{2}\left(Dc\right)=0\label{eq:ficks}
\end{equation}

If the problem is constrained to x axis, eq.~\ref{eq:ficks} can
be written as:

\begin{equation}
\frac{\partial c}{\partial t}-\frac{\partial c}{\partial x}\left(D\frac{\partial c}{\partial x}\right)=\frac{\partial c}{\partial t}-D\frac{\partial^{2}c}{\partial x^{2}}=0
\end{equation}

Here \emph{c(x,t)} represents {[}A{]} in \emph{x} position and at
\emph{t} time. The diffusion coefficient \emph{D} is assumed to be
independent of time and position, although it can show a strong temperature
(\emph{T}) dependence. Arrhenius form is commonly used to estimate
values of diffusion coefficient:

\begin{equation}
D=D_{0}\exp\left(\frac{-Q}{RT}\right)
\end{equation}

A possible solution for eq.~\ref{eq:ficks} can be defined with the
use of the error function $\textrm{erf}(z)$ as follows~\cite{Crank1975,Bliersbach2011}:

\begin{equation}
\frac{c_{s}-c_{x}}{c_{s}-c_{0}}=\textrm{erf}(\frac{x}{2\sqrt{Dt}})\label{eq:Solut}
\end{equation}

Where $c_{s}$ is the constant surface concentration of specie A,
$c_{0}$ is the initial uniform {[}A{]} into the B matrix, and $c_{x}$ is the concentration of the specie A at a certain position x.
The error function can be computed as:

\begin{equation}
\textrm{erf}(z)=\frac{2}{\sqrt{z}}\int_{0}^{z}\exp(-y^{2})dy
\end{equation}

Equation~\ref{eq:Solut} allows to estimate {[}A{]} variation in
function of time and distance, provided that adequate constants are
previously known ($D,\, c_{0},\, c_{s}$).\\ 

These equations can be solved numerically in the continuum aided by computational codes. For the purpose of the
present article, next section shows an example problem with the same
parameters that were used in our course at University of Burgos.

\section{Case problem. Hydrogen diffusion through vanadium.}

\subsection{Validation through experimental set-up.}

The use of recent innovative results 
to illustrate theory with a practical example (session 1-B), helps to put students into contact with real physics, what is positive to raise their attention and motivation.
Recent scientific achievements related to the study of diffusion in nano-sized materials, have been carried out at Uppsala University. In this case,
a new measurement technique was developed to obtain accurate results,
even with low concentrations of the diffusive specie. Traditional
methodologies, such as nuclear magnetic resonance, quasi-elastic neutron
scattering or mechanical relaxation methods, have been widely and
extensively used to study hydrogen diffusion rates.
However, they cannot offer the time-evolution profile of concentrations
inside the samples. On the contrary, an innovative technique based on an optical
setup, which can be applied on any metal, has been developed  ~\cite{Palsson2012}. A monochromatic light beam
is emitted by a a light-emitting diode (LED) source, and then split
on a semitransparent mirror in order to allow two simultaneous measurements
(intensity of source and optical transmission through the sample).
The illumination is homogenized by a diffusor before passing trough
a vacuum chamber, inside which the sample is fixed. The passing light
is finally registered by a charged coupled device (CCD), and optical
transmittance can be simultaneously monitored.

The method was employed to quantify the movement of hydrogen through
a thin film of vanadium ~\cite{Bliersbach2011}.
Samples were prepared over a monocrystalline double side polished
magnesium oxide substrate, which is transparent and have a lattice
parameter comparable with vanadium, and then covered with a layer
of polycrystalline palladium, in order to protect the sample from
oxidation. Palladium exhibits a much weaker affinity for hydrogen
than vanadium, hence concentration in palladium can be neglected.
Sample dimensions were a 100 mm x 100 mm, with 50 nm of thickness.
Vanadium is covered with aluminum zirconium which quickly is oxidized
in air to obtain a hydrogen impermeable layer, except for a window
at one of the ends. The experiment starts when molecular hydrogen
is introduced into the chamber, which is dissociated on the palladium
film. Atomic hydrogen goes inside vanadium, migrating laterally and
changing concentration profile. The previously explained apparatus
can register light penetration and convert it, recording the concentration
at short-time intervals. An example of visual output can be watched
on-line at \url{http://www.nature.com/ncomms/journal/v3/n6/extref/ncomms1897-s1.mov}.

The above described experimental workbench was decided to be used as a template to design a computer-aided simulated docent experiment.
Simulations using open-source code, aiming the understanding of theoretical concepts and expressions, where validated by means of published experimental results.

\subsection{Review of main concepts.}

After description of the experimental approach during session
1-B, objectives and concepts were reviewed to
accomplish students comprehension:
\begin{enumerate}
\item \textbf{Diffusion} mechanism explains the migration of one specie
inside another one.
\item \textbf{Fick's second law} allow performing calculations of concentration
in function of time and position. This equation requires a well known
and defined boundary conditions.
\item The key parameter for calculations is the \textbf{diffusion coefficient},
which is directly related with the two species involved. This value
can vary in function of temperature, and even position and concentration
along the samples. Average values are used in engineering procedures
to simplify the problem, but obtaining accurate enough results for
applications.
\item A \textbf{new experimental measurement technique} will be discussed putting on the table a technical report from Uppsala University ~\cite{Bliersbach2011}. Data were extracted, and compared with theoretical
expressions.
\item Finally, a \textbf{computer-based experiment} simulation will be configured
to validate the used code by statistical comparison. Students should
be able to reproduce the experiment with similar conditions, changing
initial concentrations, diffusion parameters or sample dimensions,
and being capable to discuss the obtained results.
\end{enumerate}

\subsection{Use of open-access resources for the students.}

\begin{figure}

\noindent \begin{centering}
\includegraphics[width=0.95\columnwidth]{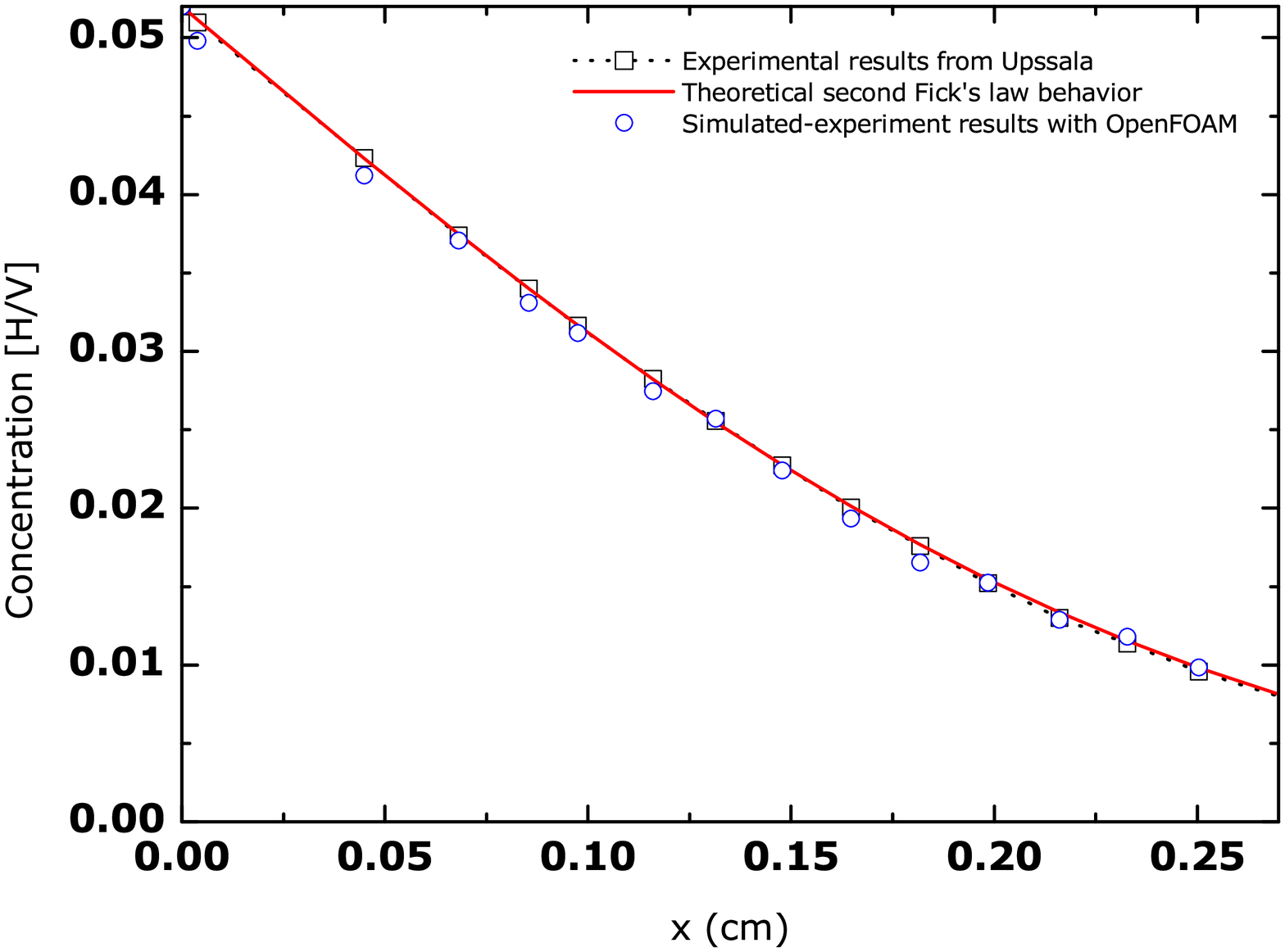}
\par\end{centering}

\caption{Hydrogen concentration profile from Bliersbach measurements as listed
in table~\ref{tab:Hydrogen-concentration-profile}, compared with
the two other approaches proposed for assessing the diffusion problem (Fick's equation,
and computer simulation using OpenFOAM\textregistered{} open-source code).\label{fig:Hydrogen-concentration-profile} }

\end{figure}

Main resources for session 1-B can be found at the on-line facilities
of Uppsala University Library, which includes a open-access repository
with scholar and research publications. The final report from Bliersbach~\cite{Bliersbach2011}
can be accessed, so descriptions about methods and diffusion results
are available for academic uses. The report declares a fitted solution
for Fick's second law which validates the proposed methodology. For
example, table~\ref{tab:Hydrogen-concentration-profile} describe
the concentration profile {[}H/V{]} for the mentioned sample geometry
at 423 {$^{\circ}$}K, and 465 s after hydrogen exposure. These data are plotted
in fig.~\ref{fig:Hydrogen-concentration-profile} showing a descent
slope. Inside Bliersbach conclusions, the average diffusivity for
a initial concentration {[}H/V{]}=0.060 and 423 {$^{\circ}$}K was computed as
$3.9\text{·}10^{-5}$m$^{2}$/s. This value is accurate enough to
be used for similar initial concentrations, as can be deduced by simple
interpolation with other data from Uppsala's report.

\begin{table}
\noindent \begin{centering}
\begin{tabular}{cc}
\toprule 
x (cm) & {[}H/V{]}\tabularnewline
\midrule
\midrule 
0.00E+00 & 0.052000\tabularnewline
\midrule 
3.88E-03 & 0.050992\tabularnewline
\midrule 
4.49E-02 & 0.042318\tabularnewline
\midrule 
6.81E-02 & 0.037395\tabularnewline
\midrule 
8.54E-02 & 0.033995\tabularnewline
\midrule 
9.74E-02 & 0.03165\tabularnewline
\midrule 
1.16E-01 & 0.02825\tabularnewline
\midrule 
1.31E-01 & 0.025553\tabularnewline
\midrule 
1.48E-01 & 0.022738\tabularnewline
\midrule 
1.65E-01 & 0.02004\tabularnewline
\midrule 
1.82E-01 & 0.017576\tabularnewline
\midrule 
1.98E-01 & 0.015229\tabularnewline
\midrule 
2.16E-01 & 0.012998\tabularnewline
\midrule 
2.33E-01 & 0.011353\tabularnewline
\midrule 
2.50E-01 & 0.00959\tabularnewline
\midrule 
2.72E-01 & 0.007826\tabularnewline
\bottomrule
\end{tabular}
\par\end{centering}

\caption{ Results extracted from Uppsala report \cite{Bliersbach2011}. Data show the hydrogen concentration profile at 423 {$^{\circ}$}K and 465~s. 
\label{tab:Hydrogen-concentration-profile}}
\end{table}

\section{Comparison between experimental measurements and theoretical behavior in our lesson.}

Regarding the docent methodology followed, Session 2 started with a brief description of the sources and goals
sought. In addition, the collaborative learning methodology was settled down with the groups formation and team leaders selection.\\
As it has been exposed previously, the solution for the eq.~\ref{eq:ficks}
gives a path to obtain the {[}A{]} at any time and position. The use
of the error function can be also substituted by the complementary
error function, obtaining a more compact expression:

\begin{equation}
c(x,t)=c_{0}\textrm{\,\ erfc}(\frac{x}{2\sqrt{Dt}})\label{eq:Solut-2}
\end{equation}

\begin{figure}

\noindent \begin{centering}
\includegraphics[width=0.95\columnwidth]{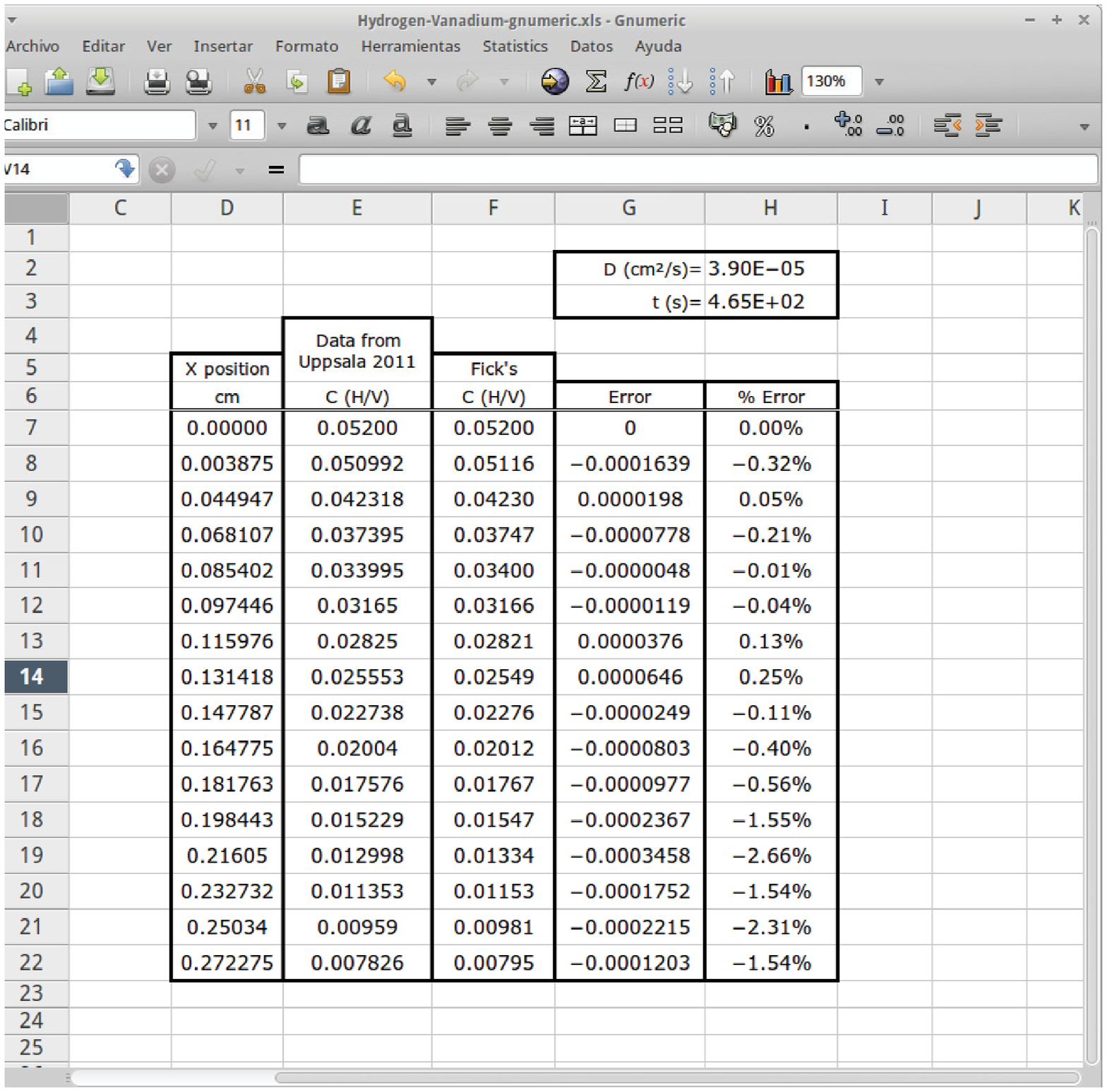}\caption{Gnumeric spreadsheet was used to quantify the initial differences
between experimental and fitted values.\label{fig:Gnumeric}}

\par\end{centering}

\end{figure}

Error function (and complementary) are available in the common spreadsheet
suites. From a figure of Upssala's reports, we can extract some useful
inputs for the practice in order to calculate the particular solution.
Table~\ref{tab:Hydrogen-concentration-profile} shows the profile
for t=465 s for an initial hydrogen concentration (t=0, x=0) {[}H/V{]}=0.052.
As mentioned above, $3.9\text{·}10^{-5}$m$^{2}$/s can be a used
as average diffusivity at 423 {$^{\circ}$}K for initial concentrations around
0.060. Therefore, session 2 includes a initial validation of published
results with the use of equations~\ref{eq:Solut} or \ref{eq:Solut-2}.
By plotting both data and error function results, an overall quantitative
comparison can be carried out (fig.~\ref{fig:Hydrogen-concentration-profile}).
Basic statistics are also a chance in this first steps. As Bliersbach
profiles (e.g., table~\ref{tab:Hydrogen-concentration-profile})
were used to fit the theoretical solutions of Fick's equation in order
to obtain a diffusion constant, negligible differences are showed
between both plots (fig.~\ref{fig:Hydrogen-concentration-profile}).
The average computed error when eq.~\ref{eq:Solut-2} is used to
predict hydrogen concentration is below 0.8 \%, with a maximum of
2.7\%. Following the idea of providing a global-availability of resources, as well as easy access to the students, the non-commercial
suite Gnumeric was selected for performing calculations, as it is showed
in fig.~\ref{fig:Gnumeric}.

\section{Running and Configuring the computer-aided simulation experiment.}

Last part of session 2 involved teaching the basis of an initial approach towards the open-source
code selected for running the simulated experiment. OpenFOAM\textregistered{}
is a validated tool for field calculations based in the Finite Volume
Method (FVM)~\cite{OpenFOAMFoundation2012}. Several solvers are
implemented in the OpenFOAM\textregistered{} main package, making
it useful for a wide range of scientific and engineering applications
(e.g., fluid dynamics, heat and mass transport, thermophysical and
chemical properties). Due to the UNIX-based structure of the code,
this part of the lessons has to be carefully step-by-step oriented, leading
to the learners comprehension of OpenFOAM\textregistered{} input requirements
to solve the problem of diffusion (table~\ref{tab:Case-struct} shows the main structure and modules needed to run this computer experiment by means of OpenFOAM\textregistered{} suite).

\begin{figure}

\noindent \begin{centering}
\includegraphics[width=0.95\columnwidth]{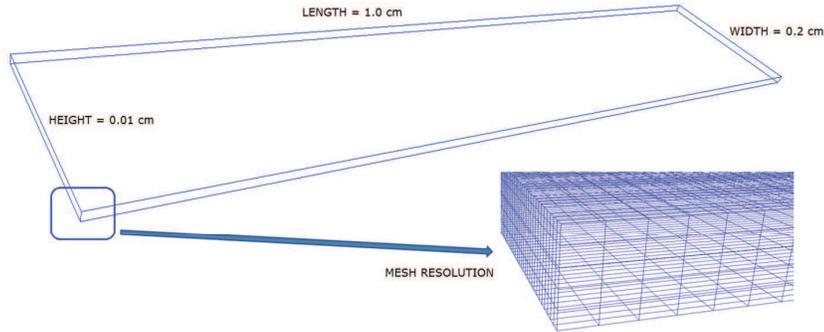}\caption{Outline of sample, and detail of mesh geometry as an input for the
simulated-experiment.\label{fig:Outline-of-sample}}

\par\end{centering}

\end{figure}

Therefore, Session 3-A must include a deep explanation of the particular case definition
(0, constant, and system folders) in OpenFOAM\textregistered{}'s framework. A standard \emph{laplacianFoam}
solver was used to design the simulated-experiment. Because of the
unidimensional nature of Uppsala observations, size of specimen can
be varied without induced errors if the total length is higher enough,
i.e. no undesirable boundary-effects are involved. Therefore, sample
dimensions were reduced in two directions to increase calculation
speed, but thickness was increased to avoid meshing problems. The
proposed shape for the simulated-experiment was as shown in fig.~\ref{fig:Outline-of-sample}.
To identify the faces, friendly names should be used for geometry
descriptions (e.g. input, output and walls). The mesh was defined
with 250x200x5 cells, and duration of simulation was initially fixed
to 500 s (storing data each 15 s).

\begin{table}
\noindent \begin{centering}
\begin{tabular}{ccc}
\toprule 
Folder & Sub-folder & Contents\tabularnewline
\midrule
\midrule 
0 &  & Boundary and Initial condition variable files\tabularnewline
\midrule 
\multirow{2}{*}{constant} &  & Other property dictionaries (transportProperties)\tabularnewline
\cmidrule{2-3} 
 & polyMesh & Mesh geometry (blockMeshDict)\tabularnewline
\midrule 
\multirow{2}{*}{system} & \multirow{2}{*}{} & Run configuration and numerical schemes\tabularnewline
 &  & (controlDict, fvSchemes; fvSolution)\tabularnewline
\bottomrule
\end{tabular}
\par\end{centering}

\caption{The basic OpenFOAM\textregistered{} case structure.\label{tab:Case-struct}}
\end{table}

Boundary conditions are key points of numeric schemes for resolving
differential equations. To simulate the same conditions used for the
fitted analytic solution, a zero gradient was imposed for temperature
over the output and lateral walls but a constant concentration was
fixed in the input ({[}H/V{]} = 0.052). The case was parallelized
before running to take advantage of multiple-core hardware.

\section{Visual postprocessing of simulation results}

One of the important features of the OpenFOAM\textregistered{} toolbox
is the 3d visualization capability, thanks to the ParaView VTK open-source
engine~\cite{Henderson2007}. A graphical output allows the learners
exploring the physical meaning of eq.~\ref{eq:ficks}. ParaFOAM is
an adapted version of ParaView, embedded as a third party software
in the main package of the tool, and configured to acquire data from
OpenFOAM\textregistered{} results.

\begin{figure}
\noindent \begin{centering}
\includegraphics[width=0.99\columnwidth]{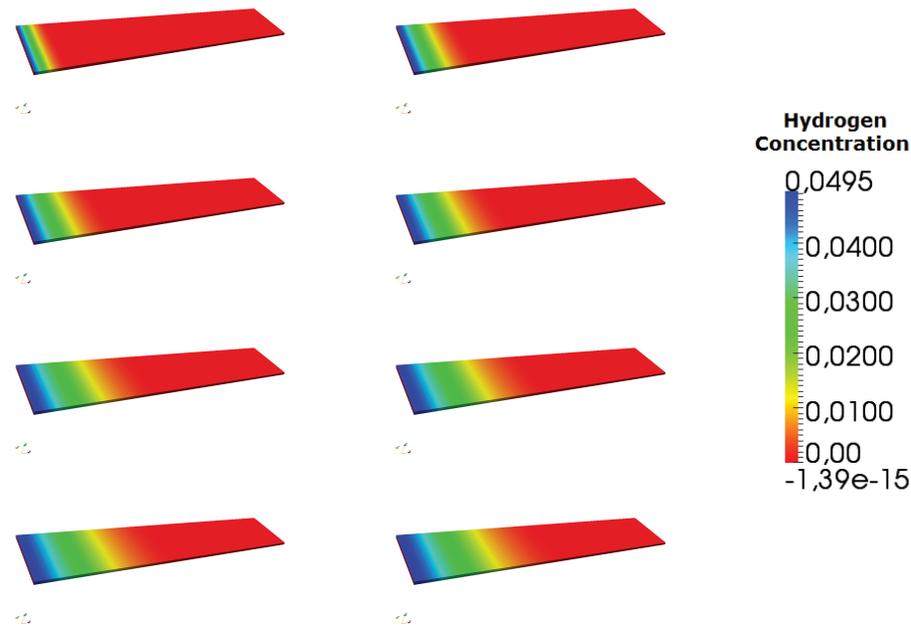}
\par\end{centering}

\caption{Time evolution of hydrogen migration inside the simulated vanadium
sheet (t= 15, 60, 105, 165, 300, 360, 420, and 465 s).\label{fig:Snapshots}}
\end{figure}

\begin{figure}
\noindent \begin{centering}
\includegraphics[width=0.95\columnwidth]{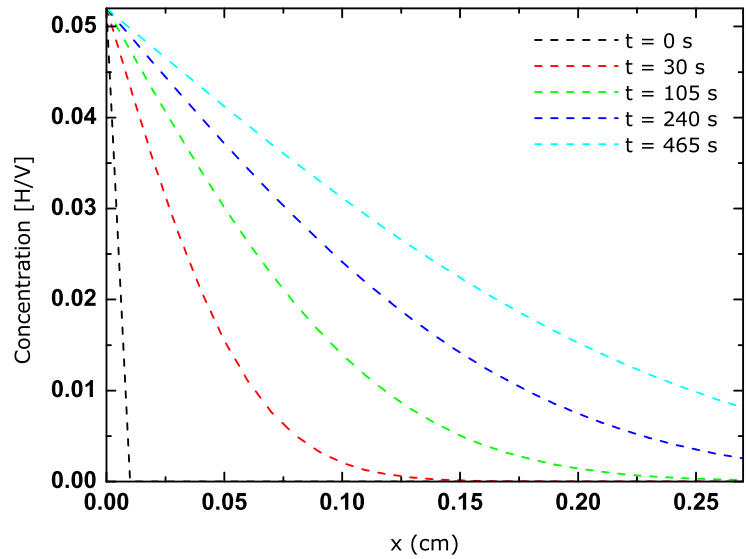}
\par\end{centering}

\caption{Profile evolution in function of distance for hydrogen concentration.
Different color lines are used for each time step, showing functions
for 0, 30, 105, 240, and 465 s.\label{fig:Profile-evolution-time}}
\end{figure}

\begin{figure}
\noindent \begin{centering}
\includegraphics[width=0.95\columnwidth]{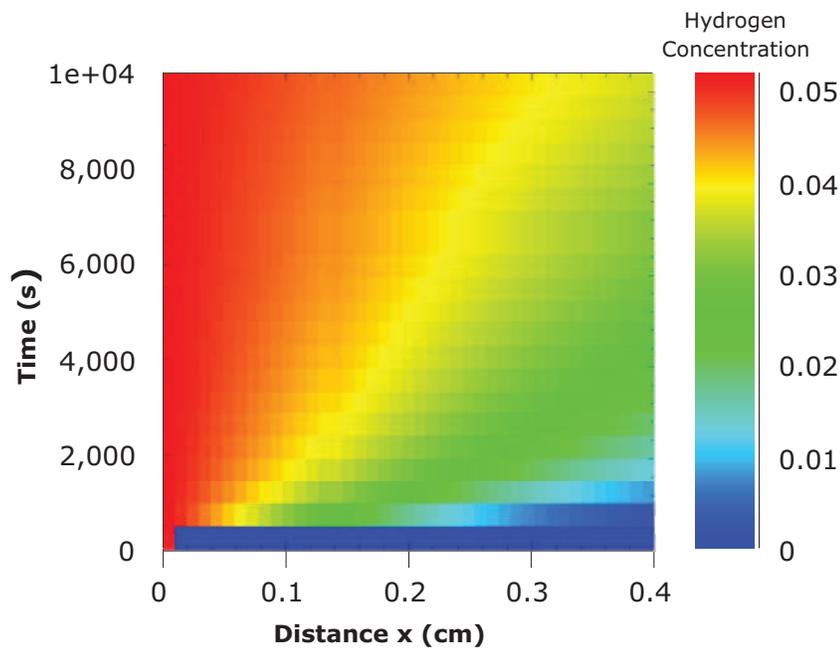}
\par\end{centering}

\caption{Contour plots of in-silico results, showing both time and position
dependence of concentration.\label{fig:Contour-plots}}
\end{figure}

The friendly interface of ParaFOAM makes easy the visualization and
analysis of concentration evolution in time scale. Examples of some
snapshots are shown in fig.~\ref{fig:Snapshots}. On the other hand,
the position dependence can be readily computed by performing different
profiles at selected times. Numerical values of these profiles can
also be extracted at screen (in a table view) and then imported to
a CSV file, which can be subsequently used at a spreadsheet to make
global charts or statistical comparisons. The time dependence of profiles
shown in fig.~\ref{fig:Profile-evolution-time} may give insight
of the atoms migration speed, which decreases when increasing time.
The experience has shown that graphical tasks are a valuable help
for students understanding. A complete contour plot can be moreover
performed (fig.~\ref{fig:Contour-plots}) to obtain another approach
to the computed solution which cannot otherwise visualized. A brief
explanation of features and capabilities of ParaFOAM was made during
the last session (3-B), giving the students a solid frame for homework
and the path for the required final report.

\section{Validation.}

A final validation was carried out among the three sets of available
data: experimental measurements of Uppsala, fitted Fick's equation
solution and in-silico simulated-experiment results. Concentration
results from OpenFOAM\textregistered{} calculations are stored in text files, including
hydrogen concentration for each one of the mesh cell-centers. Each
time-step is separated in different folders from time t=0 to t=final-step.
Students are free to open each one of concentration files, but there
is no way to recognize directly the nodes or cells involved for each
stored data. Hence, a previous visualization work is the best to obtain
coherent lists of concentrations for different time steps, identifying
concentration evolution with exact time and position parameters. Data
extraction can be constrained to a single longitudinal line due to
the expected unidimensional behavior. 

A qualitative comparison can afterwards be assessed by a overall graph,
as showed in fig.~\ref{fig:Hydrogen-concentration-profile}, which
was used to legitimate the proposed practice. Final statistics gave
an average error of 0.90 \% between simulated and laboratory experiments.

\section{Conclusions}

An open-source computational tool, OpenFOAM\textregistered{}, has been
validated for its use in a docent scheme for fluid mechanics.
Physics of non steady state diffusion of light atom species have been revisited and exported to be solved aided by an open-source computational tool.

In order to gain insight the physics of diffusion, a particular case at the forefront of research, has been studied. 
Diffusion predictions of hydrogen in vanadium were carried out comparing two different procedures:
a real published experiment that brings precise data thanks to a new innovative technique, and fitted theoretical functions. 

Our lessons helped both students and teachers to clarify the following
key concepts: Fick's second
law behavior, diffusion coefficient, importance of boundary conditions,
time and position dependence of migration specie concentration, and
the need of validation procedures for computational codes.

With a multi-session programme, a simulated-experiment practice was
performed to explain the light atoms diffusion behavior to undergraduate
engineering students. Two main objectives have been achieved: students
motivation about physics learning, and increased academic performance.
The experience has shown to the authors that better understanding
results are obtained when coupling real laboratory and simulated-experiments,
taking advantage of a graphical postprocessing 3D-suite and encouraging
students to obtain representative plots as outputs. Different proposals
of graphical examples are provided to attendees for didactic objectives,
in order to give a clear pathway for their final summary. Therefore,
the use of new computational techniques as learning tools have been
demonstrated to be a good way to draw the attention of engineering
students and reach several learning benefits.

\section*{References}{\bibliographystyle{unsrt}
\bibliography{references}
}
\end{document}